\documentclass[a4paper,10pt]{article}
\usepackage[text={7in,9in},centering]{geometry}
\usepackage[utf8]{inputenc}
\usepackage{subeqn}
\usepackage{graphicx,amssymb}
\usepackage{ulem} 
\usepackage{xcolor}
\usepackage{setspace}

\newcommand{\bk}{{\bf k}}
\newcommand{\bq}{{\bf q}}

\newcommand{\bu}{{\bf u}}
\newcommand{\bw}{{\bf w}}
\newcommand{\br}{{\bf r}}
\newcommand{\bx}{{\bf x}}
\newcommand{\by}{{\bf y}}
\newcommand{\bz}{{\bf z}}
\newcommand{\N}{{\mathbb N}}
\newcommand{\Z}{{\mathbb Z}}

\begin{document}
\title{Controlling the electron-phonon heat exchange in a metallic film by its position in a dielectric slab}

\author{
	D. V. Anghel\thanks{Institutul National de Cercetare-Dezvoltare pentru Fizica si Inginerie Nucleara Horia Hulubei, 077125 Magurele, Ilfov, Romania, \\
	Research Institute of the University of Bucharest (ICUB), 050663 Bucharest, Romania, \\
	BLTP, JINR, Dubna, Moscow region, 141980, Russia,
	dragos@theory.nipne.ro},
	M. Dolineanu\thanks{Institutul National de Cercetare-Dezvoltare pentru Fizica si Inginerie Nucleara Horia Hulubei, 077125 Magurele, Ilfov, Romania, \\
	Doctoral School of Physics, University of Bucharest, Faculty of Physics, 077125 Magurele, Ilfov, Romania,
	mircea.dolineanu@theory.nipne.ro},
	J. Bergli\thanks{Department of Physics, University of Oslo, PO Box 1048, Blindern,
		0316 Oslo, Norway, joakim.bergli@fys.uio.no},
	and I. J. Maasilta\thanks{Nanoscience Center, Department of Physics, University of Jyvaskyla, FI-40014 Jyv\"asky\"a, Finland, ilari.j.maasilta@jyu.fi}}

\date{\today}

\maketitle

\begin{abstract}
We theoretically study the heat flux between electrons and phonons in a thin metallic film embedded in a suspended dielectric slab (called a \textit{membrane}, in accordance with the established nomenclature), forming a layered structure.
The thickness of the membrane is much smaller than the other two dimensions and, in the considered temperature range, is comparable to the dominant phonon wavelength.
The thickness of the metallic layer is an order of magnitude smaller than the thickness of the membrane.
While the dependence of the heat exchange on the thicknesses of the film and of the membrane has been studied before, it is not yet known how this depends on the position of the film inside the membrane.
Here we show that the position strongly influences the heat exchange.
If we denote by $T_e$ the effective temperature of the  electrons in the metal and by $T_{ph}$ the effective temperature of the phonons (assumed to be uniform in the  entire system), then we may write in general the heat power as $P \equiv P^{(0)}(T_e) - P^{(0)}(T_{ph})$, where $P^{(0)}(T) \equiv P_s^{(0)}(T) + P_a^{(0)}(T)$, with $P_s^{(0)}(T)$ and $P_a^{(0)}(T)$ being the contributions of the symmetric and antisymmetric Lamb modes, respectively.
In the low temperature limit, we may write $P_s^{(0)}(T) \equiv C_s T^4$ and $P_a^{(0)}(T) \equiv C_a T^{3.5}$, where $C_s$ is independent of the position of the film inside the membrane, whereas $C_a$ increases with the distance between the mid-plane of the film and the mid-plane of the membrane, being zero when the film is at the center of the membrane.
Our examples show that by changing the position of the film inside the membrane one may change the electron-phonon heat power by orders of magnitude, depending on the dimensions and the temperature range.
\end{abstract}

\section{Introduction} \label{sec_intro}

Nanosystems are of great importance for current technological applications.
Therefore, understanding their physical properties is necessary for both basic science and technology development.
One such property is the electron-phonon coupling and heat exchange in nanoscopic systems consisting of metallic films in contact with dielectric suspended membranes, since structures like this appear, for example, in ultrasensitive detectors~\cite{Enss,  RevModPhys.78.217.2006.Giazotto, PhysRevApplied.16.034051, Quaranta_2013, ApplPhysLett.78.556.2001.Anghel} and microrefrigerators~\cite{RevModPhys.78.217.2006.Giazotto, RepProgrPhys.75.046501.2012.Muhonen, ApplSupercond.5.227.1998.Leivo, ApplPhysLett.70.1885.1997.Manninen, ApplPhysLett.92.163501.2008.Miller, Vercuyssen, Nguyen}.
At low temperatures, the electron-phonon heat exchange becomes weak enough that one can consider the electrons and the acoustic phonons in separate thermal equilibrium, at effective temperatures $T_e$ and $T_{ph}$, respectively.
Then, the heat exchange may be written in general as $P(T_e, T_{ph}) \equiv P^{(0)}(T_e) - P^{(1)}(T_{ph})$, but the thermal equilibrium condition $P(T, T) = 0$ implies that $P^{(0)}(T) = P^{(1)}(T)$ is the same function.
If the ``exponent'' $x \equiv d\ln[P^{(0)}(T)]/ dT = d\ln[P^{(1)}(T)]/ dT$ is constant on a wide temperature range (orders of magnitude), then one may use the approximation $P \propto T_e^x - T_{ph}^x$.
For example, in clean three-dimensional (3D) bulk systems, where the electron mean free path is longer than the thermally dominant phonon wavelength, $x = 5$~\cite{SovPhysJETP.4.173.1957.Kaganov, PhysRevLett.59.1460.1987.Allen, PhysRevB.49.5942.1994.Wellstood}, whereas for (clean, non-disordered) two-dimensional (2D) phonons in graphene, $x=4$~\cite{PhysRevB.81.245404.2010.Viljas}, and for a quasi one-dimensional (1D) phonon system $x=3$ \cite{PhysRevB.77.033401.2008.Hekking} (clean limit).
Thus, it would at first seem that$x = s+2$, where $s$ is the dimensionality of the phonon gas.

However, the above statement in not generally true, as was shown in previous theoretical studies of the electron-phonon heat exchange in thin quasi-2D suspended
layered nano-structures~\cite{SolidStateCommun.227.56.2016.Anghel, PhysRevB.93.115405.2016.Cojocaru, EurPhysJB.90.260.2017.Anghel}.
In those studies, the structure  consists of a metallic film, of a thickness of the order of 10~nm, on top of a dielectric membrane, of a thickness of the order of 100~nm.
Then, in the low temperature limit (which, for the parameters mentioned above, is of the order of 100~mK or below), the heat power flow between electrons and phonons obeys the simple power law dependence on temperature $P(T_e, T_{ph}) \propto T_e^{3.5} - T_{ph}^{3.5}$ -- so, $x = 3.5$ and $s=1.5$~\cite{SolidStateCommun.227.56.2016.Anghel, EurPhysJB.90.260.2017.Anghel}.
But as the temperature increases, $x$ starts to vary in a wide range, from 3.5 reaching  approximately 4.7 at around 0.5~K~\cite{PhysScr.94.105704.2019.Anghel}.
This is close to the experimentally observed value of $x \sim 4.5$, measured for both SiN/Cu~\cite{PhysRevLett.99.145503} and SiO2/Au \cite{Saira2020}  suspended membrane devices.

In contrast to previous work \cite{SolidStateCommun.227.56.2016.Anghel, PhysRevB.93.115405.2016.Cojocaru, EurPhysJB.90.260.2017.Anghel, PhysScr.94.105704.2019.Anghel} where the metallic film was located on top of a dielectric membrane, here we study the effect of the position of the metallic film {\em inside} the dielectric membrane on the electron-phonon heat exchange.
We observe that while in the high  temperature range (roughly above 1~K for the parameters used here) the heat exchange is almost independent of the position of the metallic film, in the low temperature sub-Kelvin limit the heat power flow decreases as the metal film is placed closer and closer to the center of the membrane, by up to one order of magnitude at 10 mK.
This provides an additional method to control the electron-phonon heat exchange, which is an important characteristic for the responsivity and noise of bolometric detectors and the effectiveness of microrefrigerators, without changing the materials or the thickness of the layers.

The article is organized as follows:
in Section~\ref{sec_methods} we describe the system and the models used, in Section~\ref{sec_results} we present the numerical results, and in Section~\ref{sec_conclusions} we draw the conclusions.

\section{Methods} \label{sec_methods}

\subsection{System description} \label{subsec_system}

\begin{figure}[b]
	\centering
	\includegraphics[width=5cm]{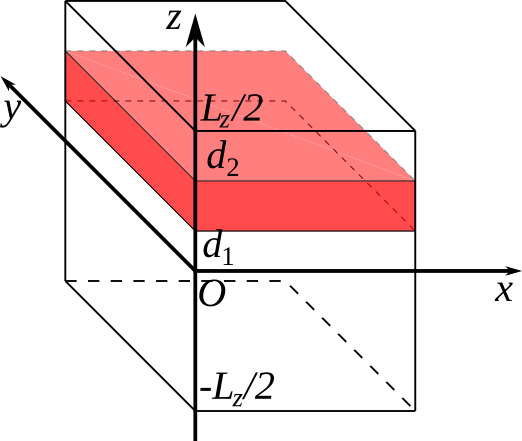}
	\caption{The structure of the system under study.
	The suspended slab is parallel to the $(x,y)$ plane and corresponds to $z \in [-L_z/2, L_z/2]$.
	The metallic layer is colored red and corresponds to $z \in [d_1, d_2]$, where $-L_z/2 \le d_1 < d_2 \le L_z/2$.
	In the $(x,y)$ plane, the system has the area $A \equiv L_x L_y$.}
	\label{fig_structure}
\end{figure}

The system, of total dimensions $L_x \times L_y \times L_z$, is schematically represented in Fig.~\ref{fig_structure} and consists of a metallic layer (red) embedded within a suspended dielectric slab.
We consider that $L_x, L_y \gg L_z$ and $L_z$ may be comparable to the dominant phonon wavelength in the temperature range of interest.
The metallic layer has the dimensions $L_x \times L_y \times d$, where $d = d_2-d_1$ is the metal layer thickness, and $-L_z/2 \le d_1 < d_2 \le L_z/2$.
Although the following equations are general, we consider in the numerical examples that $L_z$ is 100~nm and $d$ is 10~nm, which are dimension scales relevant for real devices.
We assume that the electron mean free path is longer than $d$~\cite{JApplPhys.119.085101.2016.Gall} and that the phonon mean free path is longer than $L_z$, and assume smooth interfaces and surfaces without diffusive scattering.
In the $x$ and $y$ directions the electron wavefunction $\psi$ is periodic (free motion), whereas at $z=d_1, d_2$ we assume Dirichlet boundary conditions ($\psi = 0$)--this is a good approximation for metals which have a tall potential barrier at the surface so that the electron wavefunction does not extend much outside of the metallic layer.
Then, we can write the electron wavefunction as
%
\begin{eqnarray}
\psi _{\bk_\parallel,n_z}(\br,t) &\equiv& \psi _{\bk_\parallel,k_z}(\br,t) = \phi _{k_{z}}(z)e^{i(\bk_\parallel \br_{\parallel }-\epsilon_{\bk_\parallel,n}t/\hbar )}/\sqrt{A},
\quad {\rm where} \nonumber \\
\phi _{k_{z}}(z) &=&
\left\{ \begin{array}{ll}
\sqrt{\frac{2}{d}}\sin \left[ \left( z - d_1\right) k_{z}\right] , & {\rm if} \quad z \in [d_1, d_2] ,  \\
0 , & {\rm if} \quad z \notin [d_1, d_2] ,
\end{array} \right.
\end{eqnarray}
where $\bk_\parallel$ and $k_{z}$ are the wave vector components parallel and perpendicular to the metal film, respectively.
The boundary conditions quantize the components of the wavevector to $k_x = 2\pi n_x/L_x$, $k_y = 2\pi n_y/L_y$, and $k_z = \pi n_z/L$, where $n_{x,y} \in \Z$ (integer), whereas $n_z \in \N$ (positive integer).
These quantization conditions induce a constant (but non-isotropic) density of states (DOS) in the $\bk$ space, namely, $\sigma_\bk \equiv \sigma_{k_x} \sigma_{k_y} \sigma_{k_z}$, where $\sigma_{k_x} \equiv L_x/(2\pi)$, $\sigma_{k_y} \equiv L_y/(2\pi)$, and $\sigma_{k_z} \equiv d/\pi$.
Similarly, we denote $\sigma_{\bk_\parallel} \equiv \sigma_{k_x}\sigma_{k_y}$ and since $\sigma_{k_x}, \sigma_{k_y} \gg \sigma_{k_z}$, we shall say that the states of constant $k_z$ form quasi-continuous 2D conduction bands, with a band index $n_z$.

If we denote by $m_e$ the electron's effective mass, then its energy is
\begin{equation} \label{def_el_en}
	\epsilon_{\mathbf{k}}= \frac{\hbar ^{2}k^{2}}{2m_{e}} = \frac{\hbar^{2}k_{\parallel}^{2}}{2m_{e}} + \frac{\hbar ^{2}k_{z}^{2}}{2m_{e}}
	\equiv \epsilon _{k_{\parallel},k_{z}} \equiv \epsilon _{k_{\parallel},n_{z}},
\end{equation}
where $k_{\parallel} \equiv |\bk_{\parallel}|$.
The minimum energy in the band $n_z$ is $\epsilon _{k_{\parallel} = 0,n_{z}} = \hbar ^{2}k_{z}^{2}/(2m_{e}) = (\hbar \pi n_z)^{2}/(2m_{e} d^2)$ and the difference in energy between two consecutive bands, at the same $k_\parallel$, is $\Delta\epsilon _{k_{\parallel},n_{z}} \equiv \epsilon _{k_{\parallel},n_{z}+1} - \epsilon _{k_{\parallel},n_{z}} = \hbar^2 \pi^2 (2n_z + 1) /(2m_{e} d^2)$.
We denote the Fermi energy by $\epsilon_F$ and define
\begin{equation}
n_F \equiv \left\lfloor \frac{\sqrt{2m_e\epsilon_F}}{\pi \hbar} d \right\rfloor ,
\label{def_nF}
\end{equation}
where $\lfloor x\rfloor$ is the biggest integer smaller or equal to $x$.
Then, $\epsilon _{k_{\parallel} = 0,n_{z}} \le \epsilon_F$ if and only if $n_{z} \le n_F$.
Therefore, at $T \ll \Delta\epsilon _{k_{\parallel},n_F}/k_B$ ($k_B$ is the Boltzmann constant), only the bands of $n_z \le n_F$ will be populated, plus, eventually, the band $n_z=n_F+1$, if $\epsilon_F$ is close enough to $\epsilon _{0,n_F+1}$.

To describe the phonons in our system, we assume that the whole slab (from $z = -L_z/2$ to $L_z/2$) may be treated as a homogeneous isotropic elastic material~\cite{SolidStateCommun.227.56.2016.Anghel, PhysRevB.93.115405.2016.Cojocaru, EurPhysJB.90.260.2017.Anghel, PhysScr.94.105704.2019.Anghel}.
Although a real slab would consist of different materials with differing elastic properties, our simplifying assumption is accurate enough to emphasize the qualitative features of the electron-phonon heat exchange we investigate.
The phonon modes in slabs have been studied before~\cite{Auld:book, PhysRevB.70.125425.2004.Kuhn, JPhysA.40.10429.2007.Anghel} and they differ from the phonon modes in bulk materials.
There are three types or polarizations: horizontal shear ($h$), symmetric ($s$), and antisymmetric ($a$) phonon modes (known as Lamb waves)~\cite{Auld:book}.
All these modes propagate in the direction parallel to the $(x,y)$ plane and are stationary waves along the $z$ axis.

The $h$ modes are simple transverse horizontal shear modes, with a displacement field parallel to the $(x,y)$ plane.
Their wave vector $\bq \equiv \bq_\parallel + q_{th} \hat{\bz}$ has the components parallel $\bq_\parallel \equiv q_{\parallel x} \hat{\bx} + q_{\parallel y} \hat{\by}$ and perpendicular to the membrane $q_{th} \hat{\bz}$, where $q_{\parallel x} = 2\pi \nu_x/L_x$, $q_{\parallel y} = 2\pi \nu_y/L_y$, and $q_{th} = \pi \nu_z /L$ (notice that here $th$ signifies $t=$~transverse and $h=$~horizontal shear).
The quantization conditions $\nu_x, \nu_y = \ldots, -1, 0, 1, \ldots$, and $\nu_z = 0, 1, \ldots$ are imposed by the periodic boundary conditions in the $\hat{\bx}$ and $\hat{\by}$ directions and free boundary conditions in the $\hat{\bz}$ direction~\cite{Auld:book}.
As in the case of electrons, the phonon modes with the same $\nu_z$ and any $\bq_\parallel$ form 2D bands~\cite{JPhysA.40.10429.2007.Anghel}.

The $s$ and $a$ Lamb modes, in contrast, are a superposition of transverse and longitudinal waves, with displacement fields oscillating in a plane perpendicular to the $(x,y)$ plane.
Both, the longitudinal and the transverse partial waves have the same component $\bq_\parallel$ of the wave vector parallel to the $(x,y)$ plane, whereas the components parallel to the $z$ axis, $q_l$ and $q_t$, respectively, satisfy the equation~\cite{Auld:book}
\begin{subequations} \label{syst_qtql}
\begin{equation}
	\frac{- 4q_\parallel^2  q_{l} q_{t}}{(q_\parallel^2- q_{t }^2)^2} = \left[ \frac{\tan( q_{t}L/2)} {\tan( q_{l}L/2)} \right]^{\pm 1}, \label{6}
\end{equation}
where the exponents $1$ and $-1$ on the right hand side (r.h.s) of Eq.~(\ref{6}) correspond to the symmetric ($s$) and antisymmetric ($a$) modes, respectively.
Equation~(\ref{6}) relate $q_l$ and $q_t$ for any $q_\parallel$ and for each polarization $s$ and $a$.
Another relation that has to be satisfied~\cite{Auld:book} for $q_t$ and $q_l$ is the Snell's law
\begin{equation}
	\omega_{q_\parallel} = c_{l}\sqrt{q_{l}^{2}+q_\parallel^{2}}
	= c_{t} \sqrt{q_{t}^{2}+q_\parallel^{2}} ,  \label{8}
\end{equation}
\end{subequations}
where $\omega_{q_\parallel}$ is the  angular frequency-wave vector dispersion relation of the mode. Solving Eqs~(\ref{syst_qtql}) we obtain an infinite, countable set of solutions $[q_{t,\nu_z,\sigma}(q_\parallel), q_{l,\nu_z,\sigma}(q_\parallel)]$, where $\sigma$ stands for the polarization $s$ or $a$, and $\nu_z = 0,1,\ldots$.
The components $q_{t,\nu_z,\sigma}(q_\parallel)$ and $q_{l,\nu_z,\sigma}(q_\parallel)$ take either real or imaginary values, but never complex values, with both, real and imaginary components~\cite{Auld:book}; when they are imaginary, we use the notation $q_{t,\nu_z,\sigma} \equiv i p_{t,\nu_z,\sigma}$ and $q_{l,\nu_z,\sigma} \equiv i p_{l,\nu_z,\sigma}$.

To make the notations uniform, in the following we make use of the doublets $\xi \equiv (\nu_z,\sigma)$, where $\nu_z = 0,1,\ldots$ and $\sigma = h,s,a$.
Then, the displacement fields of all the phonon modes are of the form
\begin{equation} \label{disp_field}
	\bu_{\bq_\parallel \xi}(\br, t) \equiv \frac{e^{i (\bq_\parallel \br_\parallel - \omega_{q_\parallel \xi} t)}}{2\pi} \bw_{\bq_\parallel \xi}(z) .
\end{equation}
The $z$ dependence of the displacement field of the phonon modes $\bw_{\bq_\parallel \xi}(z)$ are normalized and explicitly given, for example, in Refs.~\cite{JPhysA.40.10429.2007.Anghel, PhysScr.94.105704.2019.Anghel}.

\subsection{Electron-phonon interaction Hamiltonian} \label{subsec_elph_int}

For the electron-phonon interaction we use the deformation potential model~\cite{Ziman:book}
\begin{equation}
	\hat{H}_{\rm def} = E_a \int_{V_{el}}d^{3}\mathbf{r}\,\hat{\Psi}^{\dagger}(\mathbf{r})\hat{\Psi} (\mathbf{r})\nabla \cdot \hat{\bu}(\mathbf{r}) . \label{def_def_pot}
\end{equation}
where $\nabla \cdot \hat{\bu}(\mathbf{r})$ is the dilatation field operator, $V_{el}=A\times d$ is the volume of the metallic layer, $E_a$ is a constant, usually taken as $E_a = (2/3) \epsilon_F$~\cite{Ziman:book, SolidStateCommun.227.56.2016.Anghel, EurPhysJB.90.260.2017.Anghel, PhysScr.94.105704.2019.Anghel}, whereas the electron field annihilation and creation operators are
\begin{equation}
	\hat{\Psi} (\mathbf{r},t) = \sum_{\bk_\parallel, k_{z}} \psi_{\bk_\parallel, k_{z}}(\br, t) \hat{c}_{\bk_\parallel, k_{z}}
	\quad {\rm and} \quad
	\hat{\Psi}^{\dagger}(\mathbf{r},t) = \sum_{\mathbf{k}_{\parallel},k_{z}}\psi_{\bk_\parallel, k_{z}}^* (\br,t) \hat{c}_{\mathbf{k}_{\parallel},k_{z}}^\dagger,  \label{def_Psi}
\end{equation}
respectively.
The operators $\hat{c}_{\mathbf{k}_{\parallel },k_{z}}$ and $\hat{c}_{\bk_\parallel, k_{z}}^{\dag}$ are the electron $k$-space annihilation and creation operators on the state $\psi _{\bk_\parallel,k_z}$.
From Eq.~(\ref{disp_field}) we write the phonon field operator
\begin{eqnarray}
	\hat{\bu} (\br) &=& \sum_{\xi,\bq_\parallel} \sqrt{\frac{\hbar}{2 \rho \omega_{\bq_\parallel \xi} }}
	e^{i (\bq_\parallel \br_\parallel - i\omega_{\bq_\parallel \xi}t )}
	\left[ \hat{a}_{\bq_\parallel \xi} \bw_{\bq_\parallel \xi}(z) + \hat{a}_{-\bq_\parallel \xi}^{\dag}  \bw^*_{\bq_\parallel \xi}(z) \right] ,
	\label{displ_operator}
\end{eqnarray}
where $\hat{a}_{\bq_\parallel \xi}^{\dag}$ and $\hat{a}_{\bq_\parallel \xi}$ are the phonon creation and annihilation operators, respectively.

\subsection{Electron-phonon heat flow} \label{subsec_elph_heat}

We follow the prescription of Ref.~\cite{PhysScr.94.105704.2019.Anghel} to calculate the electron-phonon heat power flow.
We apply the Fermi golden rule to obtain from Eq.~(\ref{def_def_pot}) the transition rate $\Gamma_{i\to f} = (2\pi/\hbar) |\langle f| \hat{H}_{\rm def}| i\rangle |^2 \delta(E_f - E_i)$ between the initial ($i$) and final ($f$) state, of energies $E_i$ and $E_f$.
Using the transition rates and assuming Fermi and Bose distributions of the electrons and phonons, respectively, we calculate the heat power flow, which may be written as (Eqs.~17 of Ref.~\cite{PhysScr.94.105704.2019.Anghel})
\begin{subequations} \label{defs_P0P1}
\begin{eqnarray}
	P(T_e, T_{ph}) &\equiv& P^{(0)}(T_e) - P^{(1)}(T_e, T_{ph}) \label{17} \\
	P^{(0)} ( T_{e} ) &\equiv& P^{(0)}_s ( T_{e} ) + P^{(0)}_a ( T_{e} ) , \label{def_P0} \\
	P^{(1)} ( T_{e}, T_{ph} ) &\equiv&  P^{(1)}_s ( T_{e}, T_{ph} ) + P^{(1)}_a ( T_{e}, T_{ph} ) , \label{def_P1} \\
	P^{(0)}_\alpha ( T_{e} ) &\equiv&  \frac{4\pi}{\hbar} \sum_{\bk_\parallel \bk_\parallel', n, n'}^{\bq_\parallel, \nu} \hbar \omega _{\bq_\parallel, \alpha, \nu} |g_{\bq_\parallel, \alpha, \nu}^{n',n}|^{2}
	[f(\beta_e \epsilon_{\mathbf{k}_\parallel -\mathbf{q}_\parallel, n'}) - f(\beta_e \epsilon_{k_\parallel,n}) ]
	n(\beta_e \epsilon_{q_\parallel, \nu}) , \label{def_P0a} \\
	P^{(1)}_\alpha ( T_e, T_{ph}) &\equiv& \frac{4\pi}{\hbar} \sum_{\bk_\parallel \bk_\parallel', n, n'}^{\bq_\parallel, \nu} \hbar \omega _{\bq_\parallel, \alpha, \nu} |g_{\bq_\parallel, \alpha, \nu}^{n',n}|^{2}
	[f(\beta_e \epsilon_{\mathbf{k}_\parallel -\mathbf{q}_\parallel, n'}) - f(\beta_e \epsilon_{k_\parallel,n}) ]
	n(\beta_{ph} \epsilon_{q_\parallel, \nu}) , \label{def_P1a}
\end{eqnarray}
\end{subequations}
where $\beta_e = 1/(k_BT_e)$, $\beta_{ph} = 1/(k_BT_{ph})$, $T_e$ is the electron temperature, $T_{ph}$ is the phonon temperature, $k_B$ is Boltzmann constant, $P^{(0)}_\alpha$ and $P^{(1)}_\alpha$ are the contributions of the $\alpha$ modes, where $\alpha = s,a$ is the polarization.
Purely transverse  waves do not contribute to the electron-phonon heat exchange in our model, so the $h$ modes do not contribute to $P$ in Eq.~(\ref{17}).
Note also that the terms $P^{(0)}(T_e)$ and $P^{(1)}(T_e, T_{ph})$ are not the heat powers from electrons to phonons and from phonons to electrons, respectively, since some terms, which cancel out are not explicitly written in Eq.~(\ref{17}).
Furthermore, $\omega _{\bq_\parallel, \alpha, \nu}$ are given by Eq.~(\ref{8}), with $q_{l,\nu_z,\sigma}(q_\parallel)$ and $q_{t,\nu_z,\sigma}(q_\parallel)$ being the solutions of Eqs.~(\ref{syst_qtql}).
In Eqs.~(\ref{defs_P0P1}), we also used the notation for the coupling constant
\begin{eqnarray}
	g_{\bq_{\parallel}, \xi}^{n',n} &=& E_a N_{q_\parallel, \xi} \sqrt{\frac{\hbar }{2\rho \omega _{\bq_\parallel, \xi}}} \int_{d_1}^{d_2}\phi _{n^{\prime }}^* (z)\phi _{n}(z)
	\left[ i\bq_\parallel \cdot \bw_{\bq_\parallel, \xi}(z)+\frac{d w_{\bq_\parallel, \xi, z}(z)}{d z}\right] dz,  \label{3a}
\end{eqnarray}
where $w_{\bq_\parallel, \xi, z}$ is the component of $\bw_{\bq_\parallel, \xi}$ along the $z$ axis, and the normalization constants are
\begin{subequations} \label{defs_NsNa}
\begin{eqnarray}
	\frac{1}{N_{q_\parallel, s, \nu}^{2}} & = & A \left\{ 4|q_t|^2 q_\parallel^2 \left|\cos\left(\frac{ q_t L}{2}\right)\right|^2 \left[ \left( |q_l|^2+q_\parallel^2 \right) \frac{\sinh(p_lL)}{2 p_l} - \left( |q_l|^2-q_\parallel^2 \right) \frac{\sin(\bar{q}_lL)}{2\bar{q}_l}
	\right] \right. \nonumber \\
	& &
	+ \left| q_t^2-q_\parallel^2 \right|^2 \left|\cos\left(\frac{q_lL}{2}\right)\right| \bigg[ \left(|q_t|^2+q_\parallel^2\right) \frac{\sinh(p_tL)}{2p_t} + (|q_t|^2 - q_\parallel^2)\frac{\sin(\bar{q}_tL)}{2\bar{q}_t}\bigg]
	\nonumber \\
	& &
	\left. -4q_\parallel^2 \left|\cos\left(\frac{q_lL}{2}\right)\right|^2 \left[ p_t (|q_t|^2 + k_\parallel^2) \sinh(p_t L) - \bar{q}_t(|q_t|^2 - q_\parallel^2)\sin(\bar{q}_tL)
	\right] \right\} , \label{def_Ns} \\
	\frac{1}{N_{q_\parallel, a, \nu}^{2}} & = & A\bigg\{4 |q_t|^2 q_\parallel^2 \left|\sin\left( \frac{ q_tL}{2}\right)\right|^2 \bigg[(|q_l|^2+q_\parallel^2)\frac{\sinh(p_lL)}{2 p_l} +(|q_l|^2-q_\parallel^2)\frac{\sin(\bar{q}_lL)}{2 \bar{q}_l}\bigg]
	\nonumber \\
	& &
	+| q_t^2-q_\parallel^2|^2 \left|\sin\left(\frac{q_lL}{2}\right)\right|^2 \bigg[(|q_t|^2+q_\parallel^2)\frac{\sinh(p_tL)}{2p_t} - (|q_t|^2 - q_\parallel^2)\frac{\sin(\bar{q}_tL)}{2\bar{q}_t}\bigg]
	\nonumber \\
	& &
	- 4 q_\parallel^2 \left|\sin\left(\frac{q_lL}{2}\right) \right|^2 \left[ p_t(|q_t|^2+q_\parallel^2)\sinh(p_tL) + \bar{q}_t (|q_t|^2 - q_\parallel^2) \sin(\bar{q}_tL) \right] \bigg \} . \label{def_Na}
\end{eqnarray}
\end{subequations}
In Eqs.~(\ref{defs_NsNa}) $\bar{q}_t$ and $\bar{q}_l$ are the real and parts of $q_t$ and $q_l$, respectively.
Since $q_l$ and $q_t$ may be either real or imaginary, the expressions~(\ref{defs_NsNa}) should be interpreted as a limit, when the redundant component goes to zero: $\lim_{p_{t/l}\to 0} \sinh(p_{t/l}L)/(2p_{t/l}) = L/2$ and  $\lim_{\bar{q}_{t/l}\to 0} \sin(\bar{q}_{t/l}L)/(2\bar{q}_{t/l}) = L/2$.
Combining Eqs.~(\ref{defs_P0P1}), (\ref{3a}) and (\ref{defs_NsNa}) we obtain (see, for example,~\cite{PhysScr.94.105704.2019.Anghel})
\begin{subequations} \label{Psa_ep_delta_exp}

	\begin{eqnarray}
		P_s^{(0)} &=& \frac{4 A}{\pi^2 L} \frac{E_a^2}{\rho c_l^4} \frac{2m}{\hbar^2} \sum_{n} \sum_{n'} \sum_{\nu} \int_0^\infty dx_\parallel \, x_\parallel \frac{I_{s,\nu}^{(0)} (x_\parallel)}{2x_\parallel}
		n(\beta_e \hbar\omega_{s,\nu,q_\parallel}) I_P , \label{Ps_ep_delta_exp} \\
		P_a^{(0)} &=& \frac{4 A}{\pi^2 L} \frac{E_a^2}{\rho c_l^4} \frac{2m}{\hbar^2} \sum_{n} \sum_{n'} \sum_{\nu} \int_0^\infty dx_\parallel \, x_\parallel  \frac{I_{a,\nu}^{(0)} (x_\parallel)}{2x_\parallel}
		n(\beta_e \hbar\omega_{a,\nu,q_\parallel}) I_P
		, \label{Pa_ep_delta_exp}
	\end{eqnarray}
\end{subequations}
where we use the dimensionless notations $y_\parallel \equiv (L/2) k_\parallel$, $x_\parallel \equiv (L/2) q_\parallel$, 
$x_{l,\xi} \equiv x_{l,\xi}(q_\parallel) \equiv q_{l,\xi}(q_\parallel) (L/2)$,
$x_{t,\xi} \equiv x_{t,\xi}(q_\parallel) \equiv q_{t,\xi}(q_\parallel) (L/2)$,
$\bar{x}_{l,\xi} \equiv \bar{x}_{l,\xi}(q_\parallel) \equiv \bar{q}_{l,\xi}(q_\parallel) (L/2)$,
$\bar{x}_{t,\xi} \equiv \bar{x}_{t,\xi}(q_\parallel) \equiv \bar{q}_{t,\xi}(q_\parallel) (L/2)$,
$\chi_{l,\xi} \equiv \chi_{l,\xi}(q_\parallel) \equiv p_{l,\xi}(q_\parallel) (L/2)$,
$\chi_{t,\xi} \equiv \chi_{t,\xi}(q_\parallel) \equiv p_{t,\xi}(q_\parallel) (L/2)$,
%
$z_\parallel \equiv \beta_e (\hbar^2/2m)(2/L)^2 y_\parallel^2$, $z_1 \equiv \beta_e (\hbar^2/2m)(n\pi/L)^2$,  $z_{\rm min} \equiv \beta_e (\hbar^2/2m)(2/L)^2 y_{\rm min}^2$, $z_{ph} \equiv \beta_e \hbar\omega_{\xi,q_\parallel}$, and $z_{\epsilon_F} \equiv \beta_e {\epsilon_F}$, and
\begin{eqnarray}
	y_{\rm min} &\equiv& \frac{L}{4 q_\parallel} \left| \frac{2m}{\hbar^2} \hbar\omega_{\xi,q_\parallel} + q_\parallel^2 + \left({n'}^2 - n^2\right) \left( \frac{\pi}{L} \right)^2 \right|
	. \label{def_ymin}
\end{eqnarray}
In Eqs.~(\ref{Psa_ep_delta_exp}) we have the integral
\begin{subequations}\label{defs_ints}
\begin{equation}
	I_P = \frac{1}{2} \sqrt{k_BT \frac{2m}{\hbar^2}} \frac{L}{2} \int_0^\infty \frac{dz'_\parallel}{\sqrt{z'_\parallel}} \left\{ \frac{1}{e^{z'_\parallel - ( z_{\epsilon_F} + z_{ph} - z_1 - z_{\rm min}) } + 1} - \frac{1}{e^{z'_\parallel - (z_{\epsilon_F} - z_1 - z_{\rm min}) } + 1} \right\} \label{IP1}
\end{equation}
and the notations
\begin{eqnarray}
	I_{s,\nu}^{(0)} (x_\parallel) &=& \sum_{n} \sum_{n'} |x_t|^2 x_\parallel^2 |\cos(x_t)|^2 |G_{s, \nu, q_\parallel}(n, n')|^2 \hbar \omega_{s, \nu,q_\parallel}^4 \nonumber \\
	&& \times \left\{ 4 |x_t|^2 x_\parallel^2 |\cos(x_t)|^2
	\left[ (|x_l|^2+x_\parallel^2) \frac{\sinh(2 \chi_l)}{2 \chi_l} + (x_\parallel^2 - |x_l|^2) \frac{\sin(2\bar{x}_l)}{2 \bar{x}_l}\right]\right.\nonumber\\
	&& + |x_\parallel^2 - x_t^2|^2 |\cos(x_l)|^2
	\left[ (|x_t|^2+x_\parallel^2) \frac{\sinh(2 \chi_t)}{2 \chi_t} - (x_\parallel^2 - |x_t|^2) \frac{\sin(2\bar{x}_t)}{2 \bar{x}_t} \right] \nonumber\\
	&& \left. - 4x_\parallel^2 |\cos(x_l)|^2 \left[ \chi_t(|x_t|^2+x_\parallel^2) \sinh(2 \chi_t) + x_t(x_\parallel^2 - |x_t|^2) \sin(2\bar{x}_t) \right] \right\}^{-1}
	, \label{Ps_ep_membrI} \\
	I_{a,\nu}^{(0)} (x_\parallel) &=& \sum_{n} \sum_{n'} |x_t|^2 x_\parallel^3 |\sin(x_t)|^2 |G_{a, \nu, q_\parallel}(n,n')|^2 \hbar \omega^4_{a, \nu,q_\parallel} \nonumber \\
	&& \times \left\{
	4|x_t|^2 x_\parallel^2 |\sin(x_t)|^2\left(
	(|x_l|^2+q_\parallel^2)\frac{\sinh(2 \chi_l)}{2 \chi_l}
	+(|x_l|^2-x_\parallel^2)\frac{\sin(2\bar{x}_l)}{2\bar{x}_l}\right)\right.\nonumber\\
	&& +|x_t^2-x_\parallel^2|^2 |\sin(x_l)|^2\left(
	(|x_t|^2+x_\parallel^2)\frac{\sinh(2 \chi_t)}{2 \chi_t}
	-(|x_t|^2-x_\parallel^2)\frac{\sin(2\bar{x}_t)}{2\bar{x}_t}\right)\nonumber\\
	&& -4x_\parallel^2 |\sin(x_l)|^2\left( \chi_t(|x_t|^2+x_\parallel^2)
	\sinh(2 \chi_t)+\bar{x}_t(|x_t|^2-x_\parallel^2)\sin(2\bar{x}_t)\right)
	\left.\vphantom{\frac{1}{2}}\right\}^{-1}
	, \label{Pa_ep_membrI}
\end{eqnarray}
\end{subequations}
with
\begin{subequations} \label{def_Gs2_Ga2_int}
\begin{eqnarray}
G_{q_\parallel, s, \nu}(n,n') &=& \frac{2}{d} \int^{d_2}_{d_1} dz \sin\left[\left(z-d_1\right)\frac{n\pi}{d}\right] \sin\left[\left(z-d_1\right)\frac{n'\pi}{d}\right] \cos[{q}_{l, s, \nu}(q_\parallel) z] , \nonumber \\
&=& - \frac{8 {\pi}^{2} n_1 n_2 x_l }
{ \left[ \pi^2 ( n_1 - n_2 )^2 \left( \frac{L}{d_2 - d_1} \right)^2 - 4 x_l^2 \right]
\left[ \pi^2 ( n_1 + n_2 )^2 \left( \frac{L}{d_2 - d_1} \right)^2 - 4 x_l^2 \right] } \nonumber \\
&& \times \left[ ( -1 )^{n_1 + n_2} \sin \left( \frac {2 x_l d_2}{L} \right) -\sin \left( {\frac {2 x_l d_1}{L}} \right) \right] \left( \frac{L}{d_2-d_1} \right)^3  \label{def_Gs2_int} \\
G_{q_\parallel, a, \nu}(n,n') &=& \frac{2}{d} \int^{d_2}_{d_1} dz \sin\left[\left(z-d_1\right)\frac{n\pi}{d}\right] \sin\left[\left(z-d_1\right)\frac{n'\pi}{d}\right] \sin[{q}_{l, a, \nu}(q_\parallel) z] . \nonumber \\
&=& \frac{8 {\pi}^{2} n_1 n_2 x_l }
{ \left[ \pi^2 ( n_1 - n_2 )^2 \left( \frac{L}{d_2 - d_1} \right)^2 - 4 x_l^2 \right]
\left[ \pi^2 ( n_1 + n_2 )^2 \left( \frac{L}{d_2 - d_1} \right)^2 - 4 x_l^2 \right] } \nonumber\\
&& \times \left[ (-1)^{n_1 + n_2} \cos \left( 2 \frac {x_l d_2}{L} \right)
- \cos \left( 2 \frac {x_l d_1}{L} \right) \right] \left( \frac{L}{d_2 - d_1} \right)^3 \label{def_Ga_int1_int}
\end{eqnarray}
\end{subequations}
In general, $z_{\epsilon_F} - z_1 - z_{\rm min} \gg 1$~\cite{SolidStateCommun.227.56.2016.Anghel, EurPhysJB.90.260.2017.Anghel, PhysScr.94.105704.2019.Anghel}, so we can use the approximation (see the Appendix of Ref.~\cite{EurPhysJB.90.260.2017.Anghel})
\begin{eqnarray}
	I_P &\approx& \frac{1}{2} \sqrt{k_BT_e \frac{2m}{\hbar^2}} \frac{L}{2} \frac{z_{ph}}{\sqrt{z_{\epsilon_F} - z_1 - z_{\rm min}}}
	= \sqrt{\frac{2m}{\hbar^2}} \frac{L}{4} \frac{\hbar\omega_{\xi,q_\parallel}}{\sqrt{{\epsilon_F} - \frac{\hbar^2}{2m} \left(\frac{2}{L}\right)^2 \left[ \left(\frac{n\pi}{2}\right)^2 - y_{\rm min}^2 \right]}}
	\label{IP1_approx}
\end{eqnarray}
and observe that $I_P$ does not depend on temperature.
In such a case, the only temperature dependence in the expressions~(\ref{Psa_ep_delta_exp}) is in the phonon populations $n(\beta_e \hbar\omega_{\sigma,\nu,q_\parallel})$.

The term $P^{(1)}$ may be calculated similarly as $P^{(0)}$, but replacing $T_e$ by $T_{ph}$ in the phonon populations $n(\beta_{ph} \epsilon_{q_\parallel, \nu})$ of Eq.~(\ref{def_P1a}), as shown in detail in Refs.~\cite{SolidStateCommun.227.56.2016.Anghel, EurPhysJB.90.260.2017.Anghel, PhysScr.94.105704.2019.Anghel}.
Therefore, in the limit $z_{\epsilon_F} - z_1 - z_{\rm min} \gg 1$, $P^{(1)}_\alpha ( T_e, T_{ph})$ remains only a function of $T_{ph}$, in such a way that we may write in general
\begin{equation}
	P^{(1)}_s(T) = P^{(0)}_s(T), \quad P^{(1)}_a(T) = P^{(0)}_a(T), \quad {\rm so} \quad P^{(1)}(T) = P^{(0)}(T) .
	\label{equal_powers}
\end{equation}
Notice that these simplifications are valid only outside the very narrow \textit{crest regions}, as they are defined in~\cite{SolidStateCommun.227.56.2016.Anghel, EurPhysJB.90.260.2017.Anghel, PhysScr.94.105704.2019.Anghel}.
Therefore, in the region of applicability of Eq.~(\ref{equal_powers}), Eq.~(\ref{17}) simplifies to
\begin{equation}
	P(T_e, T_{ph}) = P^{(0)}(T_e) - P^{(0)}(T_{ph}) .
	\label{17_v2}
\end{equation}
In the low temperature limit, only the lowest phonon band is populated and the expressions~(\ref{def_Gs2_Ga2_int}) are simplified to
\begin{subequations} \label{def_Gs2_Ga2_int_lT}
\begin{eqnarray}
	G_{q_\parallel, s, \nu}(n,n') &=&
	\left\{ \begin{array}{ll}
		1 , & {\rm if} \quad n - n' = 0 , \\
		4 \chi_l^2 \frac{(d_2-d_1)^2}{\pi^2 L^2} \bigg\{ \frac{ 1 }{(n-n')^2} - \frac{1}{(n+n')^2} \bigg\}
		, & {\rm if} \quad n - n' = 2 k , \\
		- \frac{ 4 \chi_l^2 (d_1+d_2) (d_2-d_1)}{\pi^2 L^2} \bigg\{ \frac{ 1 }{(n-n')^2} -  \frac{ 1 }{(n+n')^2} \bigg\}
		, & {\rm if} \quad n - n' = 2k+1 ,
	\end{array} \right.
	\label{def_Gs2_int_lT} \\
	G_{q_\parallel, a, \nu}(n,n') &=&
	\left\{ \begin{array}{ll}
		i \chi_l \frac{d_2 + d_1}{L} , & {\rm if} \quad n - n' = 0 , \\
		\frac{4}{\pi^2} i \chi_l^3 \frac{(d_2 + d_1) (d_2-d_1)^2}{L^3} \bigg\{ \frac{ 1 }{(n-n')^2} - \frac{ 1 }{(n+n')^2} \bigg\}
		, & {\rm if} \quad n - n = 2k , \\
		-\frac{4}{\pi^2} i\chi_l \frac{d_2-d_1}{L} \bigg\{ \frac{ 1 }{(n-n')^2} - \frac{ 1 }{(n+n')^2} \bigg\}
		, & {\rm if} \quad n - n' = 2k + 1 .
	\end{array}\right.
	\label{def_Ga_int1_int_lT}
\end{eqnarray}
\end{subequations}
The main contribution to the heat power flow (especially in the low temperature limit) comes from the cases $n=n'$, since in the other cases the heat exchange involve phonons of very high energy (for a typical 10~nm thick Cu metallic film, the lowest energy difference between two bands is $\Delta \epsilon_{k_\parallel=0, n_z = 1} \approx 131$~K~\cite{EurPhysJB.90.260.2017.Anghel}).

From Eq.~(\ref{def_Gs2_int_lT}) we observe that in the low temperature limit $G_{q_\parallel, s, \nu}(n,n)$, and therefore $P_s^{(0)}$, is independent of the position of the metallic layer in the dielectric membrane.
On the other hand, from Eq.~(\ref{def_Ga_int1_int}) we notice that $G_{q_\parallel, a, \nu}(n,n) = 0$ for $d_1=-d_2$ (when the metallic film is in the middle of the membrane).

\section{Results} \label{sec_results}

\begin{figure}[t]
	\centering
	\includegraphics[width=55mm]{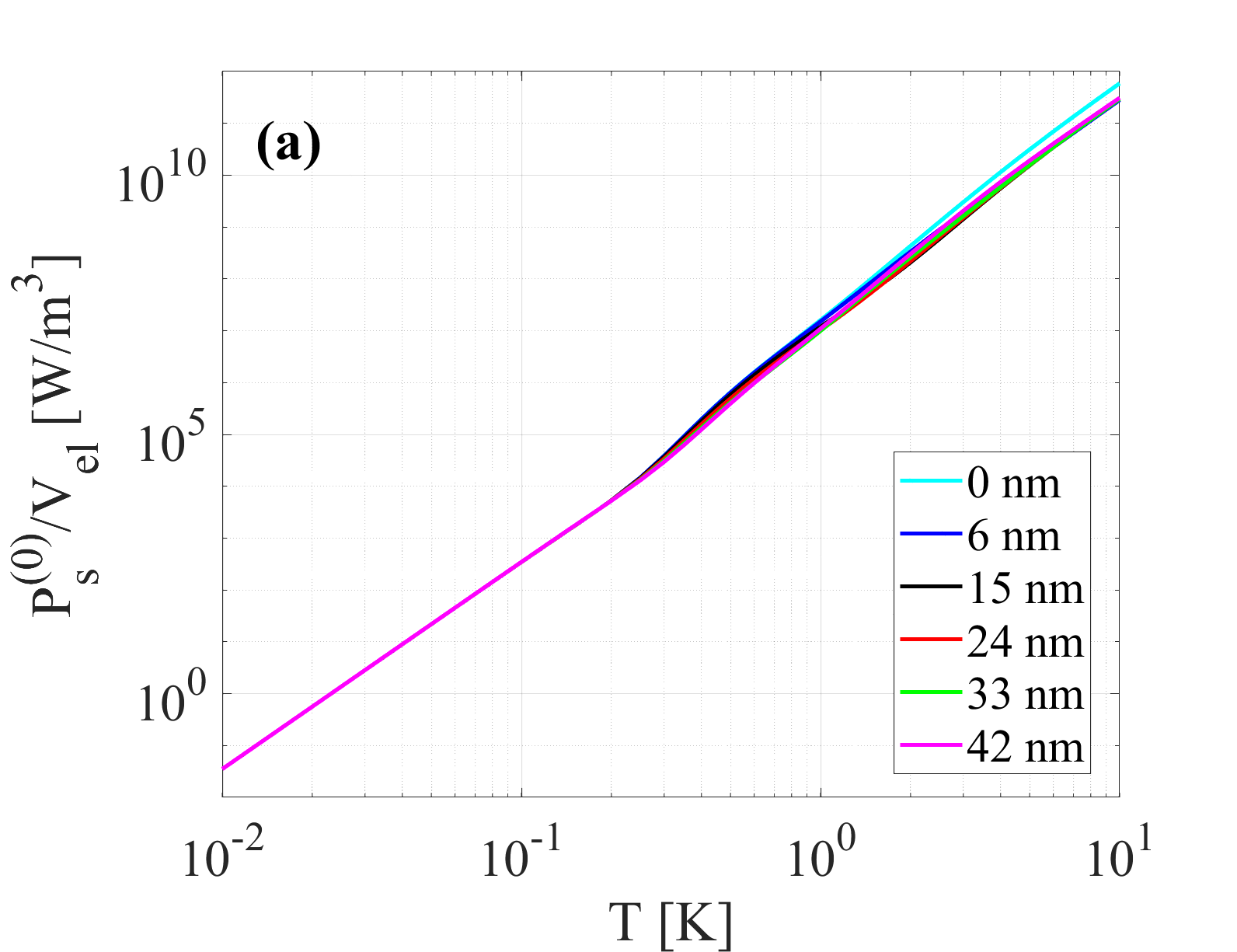}
	\includegraphics[width=55mm]{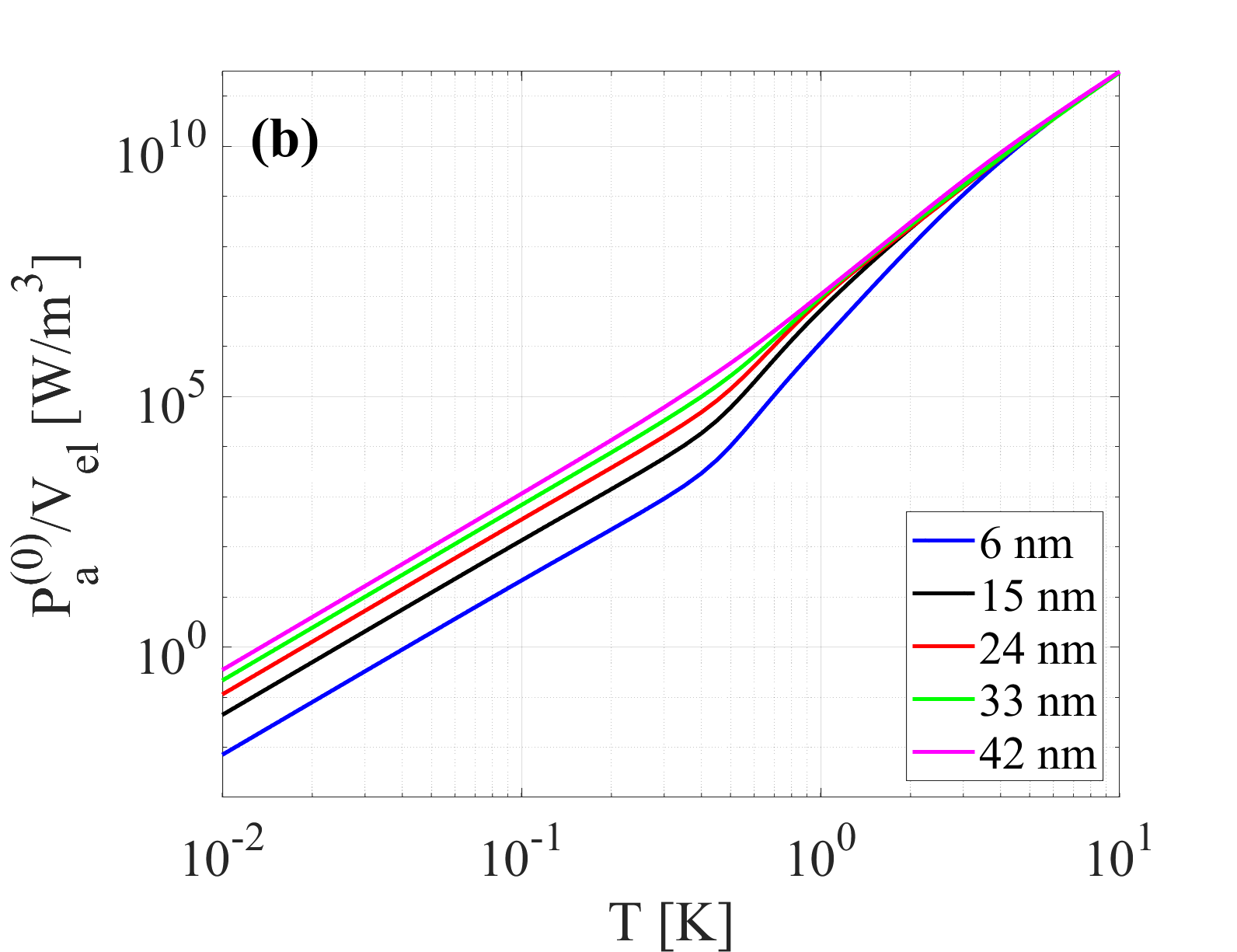}
	\includegraphics[width=55mm]{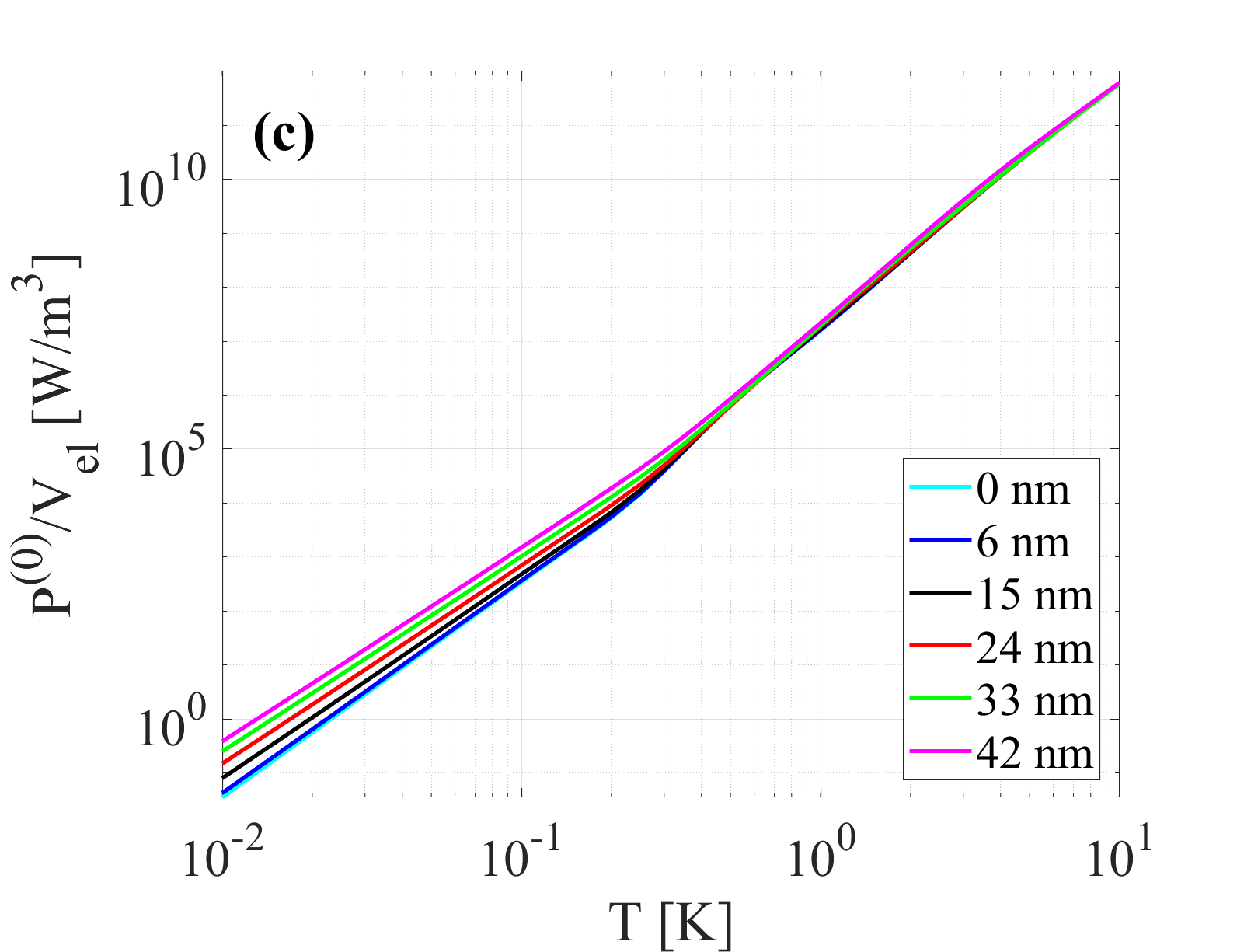}
	\caption{The heat power flow density between electrons and phonons: (a) the contribution of the symmetric Lamb-modes, (b) the contribution of the antisymmetric Lamb-modes, and (c) the total power density.
	The materials considered are SiN and Cu, and the system dimensions are $L=100$~nm and $d=d_2-d_1=10$~nm.
	The legends indicate the position of the middle of the metallic film, $(d_1+d_2)/2$.}
	\label{fig_p0s}
\end{figure}

\begin{figure}[b]
	\centering
	\includegraphics[height=45mm]{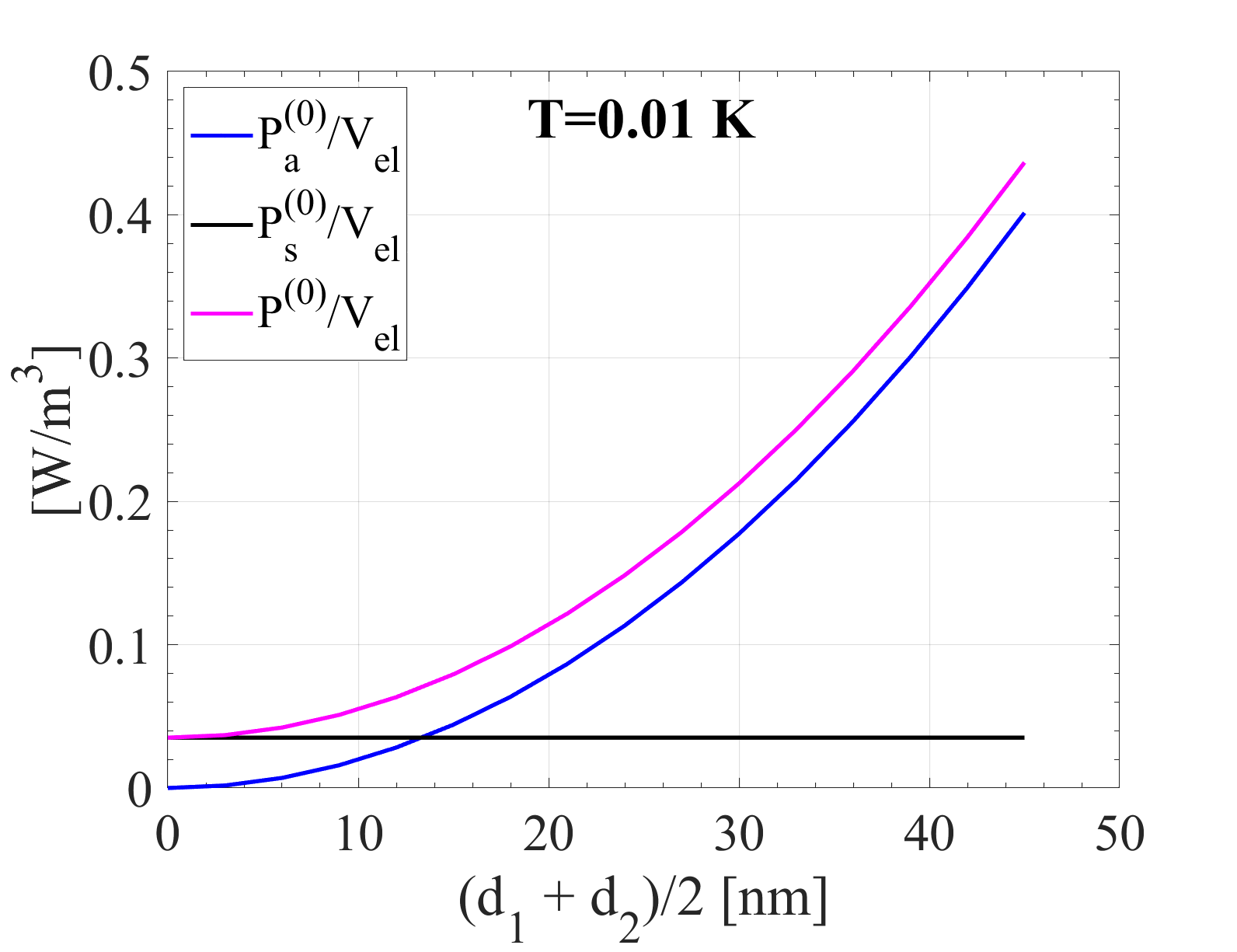}
	\includegraphics[height=45mm]{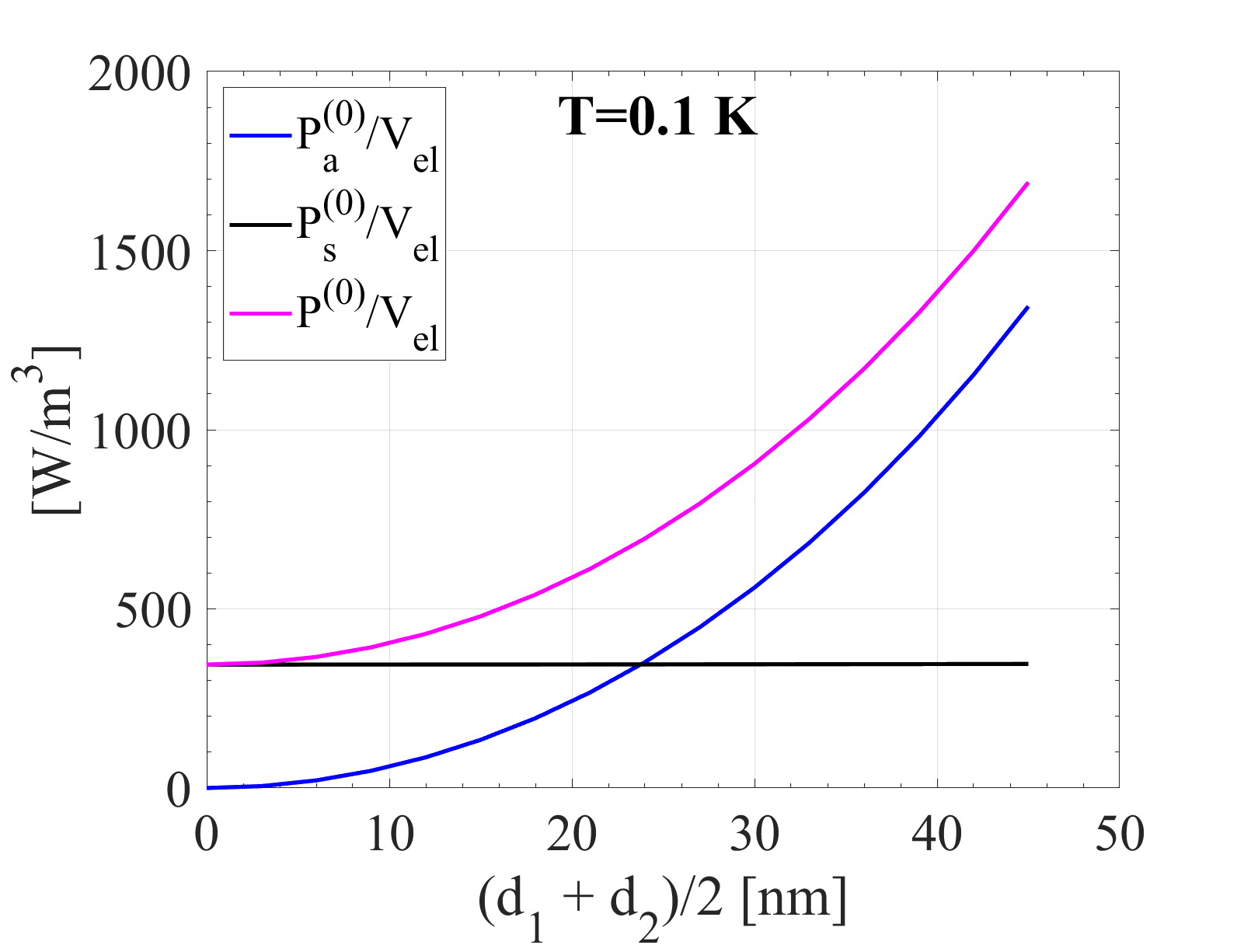}
	\includegraphics[height=45mm]{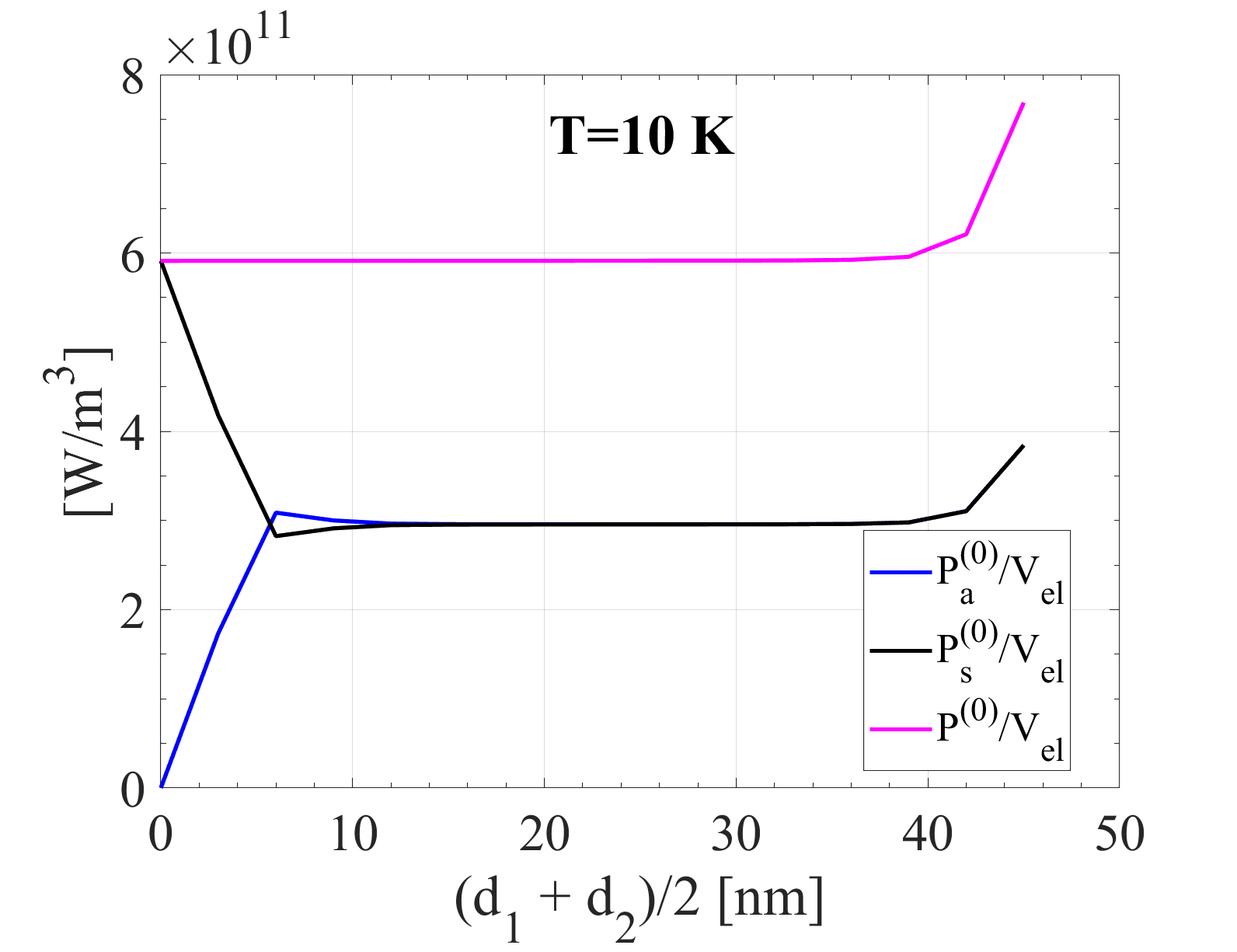}
	\caption{The heat power flow density between electrons and phonons as a function of the position of the metallic Cu film (10~nm thick) inside the dielectric SiN membrane (100~nm thick), for three temperatures: $T=0.01, 0.1, 10$~K; magenta line = total, black line = $s$-mode, blue line = $a$-mode contribution.}
	\label{fig_t001}
\end{figure}

Let us consider a 10~nm thick Cu film at an arbitrary location inside a 100~nm thick suspended SiN$_x$ dielectric slab.
The density of SiN$_x$ is 3290~kg/m$^3$, whereas the longitudinal and transversal sound velocities are 10300~m/s and 6200~m/s, respectively.
The Fermi energy in Cu is 7~eV and the 10~nm thick Cu film is outside the \textit{crest region}~\cite{SolidStateCommun.227.56.2016.Anghel, EurPhysJB.90.260.2017.Anghel, PhysScr.94.105704.2019.Anghel}, so we can use the expression~(\ref{IP1_approx}) for $I_P$.
In the temperature range of interest (from 10~mK to 10~K) we can use only the terms $n=n'$ in the summations~(\ref{Ps_ep_membrI}) and (\ref{Pa_ep_membrI}).
In Fig.~\ref{fig_p0s} we plot $P_s^{(0)}$, $P_a^{(0)}$, and $P^{(0)} = P_s^{(0)} + P_a^{(0)}$ as functions of $T$, for different positions of the Cu film in the membrane, specified by $(d_1+d_2)/2$.
We notice that in the low temperature range (say, around 100~mK and below) $P_s^{(0)}$ is practically independent of the position of the film, confirming Eq.~(\ref{def_Gs2_int_lT}), whereas $P_a^{(0)}$ strongly depends on it in the whole temperature range investigated, giving no contribution when the film is exactly in the middle, $P_a^{(0)} = 0$ at $d_1=-d_2$.
This can be seen more clearly in Fig.~\ref{fig_t001}, where we plot $P_s^{(0)}$, $P_a^{(0)}$, and $P^{(0)}$ as functions of the Cu film position $(d_1+d_2)/2$ at three different temperatures: $T=0.01$~K, $T=0.1$~K, and $T=10$~K.
We notice that in the sub-K temperature range, there is a crossover from the symmetric-mode domination for close-to-central metal film locations,  to the antisymmetric-mode domination in the opposite limit.

As it was noticed also in Refs.~\cite{SolidStateCommun.227.56.2016.Anghel, EurPhysJB.90.260.2017.Anghel, PhysScr.94.105704.2019.Anghel}, in the low temperature limit, $P_s^{(0)}$ decreases faster than $P_a^{(0)}$ with decreasing temperature, so, for any $d_1+d_2 \ne 0$, there is a crossover temperature $T_c(d_1+d_2)$, such that $P_s^{(0)} < P_a^{(0)}$ for $T<T_c(d_1+d_2)$.
Therefore, at low enough temperatures, the heat power exchanged by the electrons with the antisymmetric phonons dominates the heat power exchanged with the symmetric phonons at any $|d_1+d_2|/2 > 0$.
Due to this variation of $P^{(0)}_a$ with the position of the metallic film, at $T=10$~mK the total heat exchange power $P^{(0)}$ decreases by as much as an order of magnitude when moving the metallic film from the surface of the slab to the middle of it.

\begin{figure}[t]
	\centering
	\includegraphics[width=55mm]{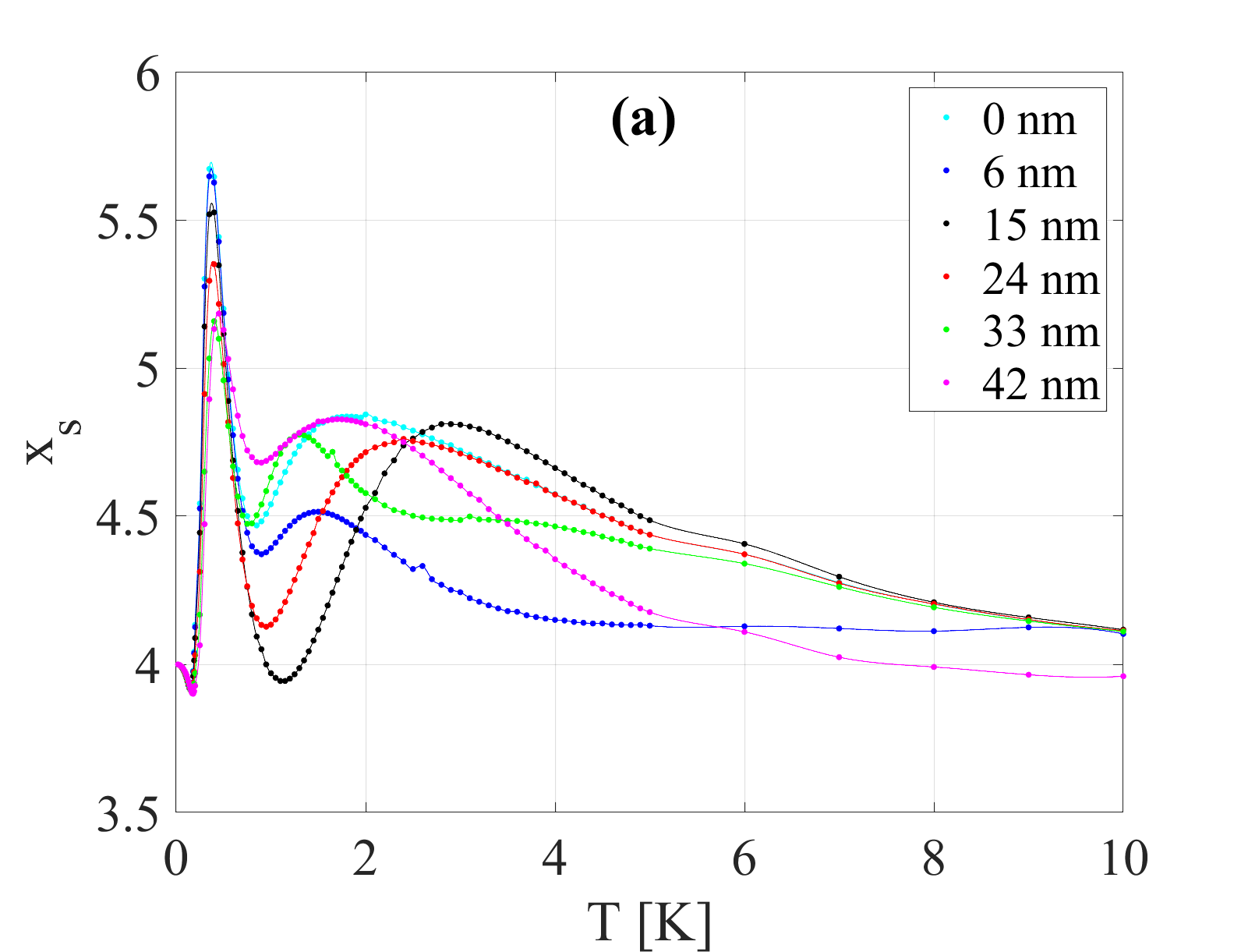}
	\includegraphics[width=55mm]{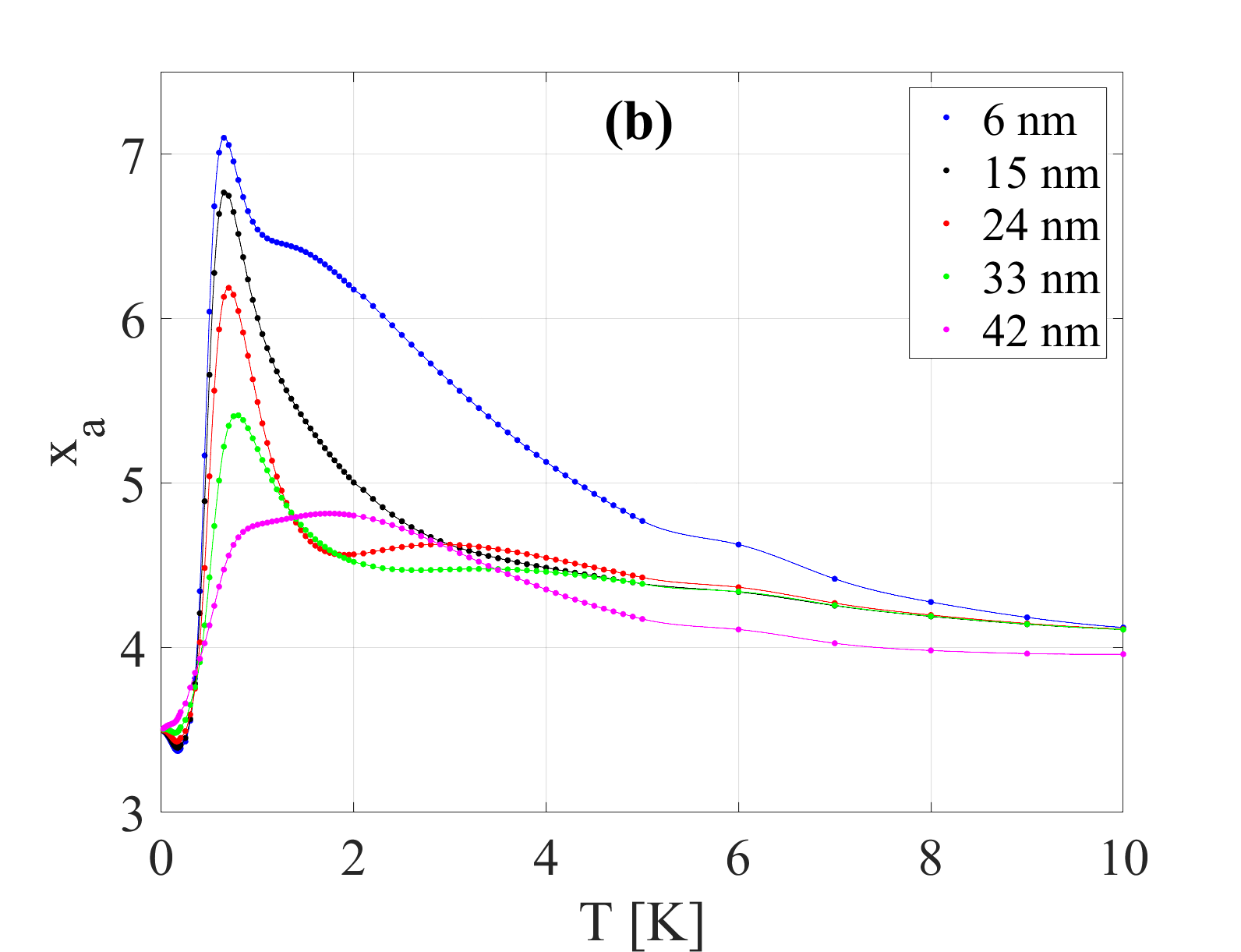}
	\includegraphics[width=55mm]{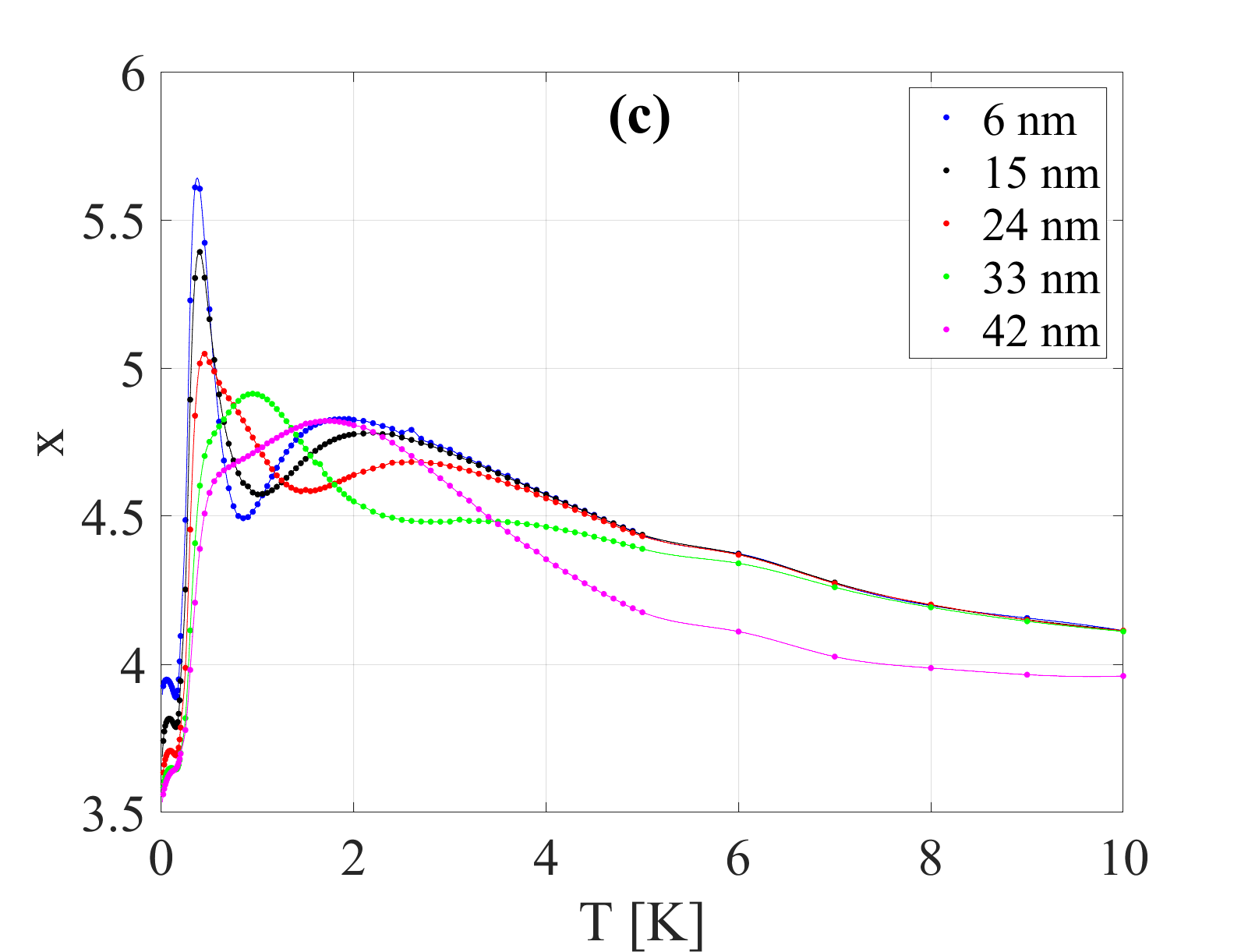}
	\caption{The exponents of the temperature dependence of the heat power flows, Eq.~\ref{defs_xsxax}, between electrons and (a) symmetric Lamb-modes, (b) antisymmetric Lamb-modes, and (c) symmetric plus antisymmetric Lamb-modes.
	The system's dimensions are $L=100$~nm (the SiN$_x$ membrane) and $d=d_2-d_1=10$~nm (the Cu film placed inside the membrane).
	The legends indicate the distance between the middle of the metallic Cu film, $(d_1+d_2)/2$, and the middle plane of the membrane}.
	\label{fig_xs}
\end{figure}

In addition, in Fig.~\ref{fig_xs} we plot the exponent of the temperature dependence for the different components of the heat power flow, defined as
\begin{equation}
	x_s \equiv \frac{\partial \ln P_s^{(0)}(T)}{\partial \ln T}, \qquad
	x_a \equiv \frac{\partial \ln P_a^{(0)}(T)}{\partial \ln T}, \qquad {\rm and} \qquad
	x \equiv \frac{\partial \ln P^{(0)}(T)}{\partial \ln T} .
	\label{defs_xsxax}
\end{equation}
One can show that~\cite{SolidStateCommun.227.56.2016.Anghel, EurPhysJB.90.260.2017.Anghel, PhysScr.94.105704.2019.Anghel}
\begin{equation}
	\lim_{T\to 0} x_s = 4 \quad {\rm and} \quad \lim_{T\to 0} x_a = 3.5 ,
	\label{lims_PsPa}
\end{equation}
so, at low enough temperatures $x_s>x_a$, as mentioned above.

At higher temperatures, the exponents $x_s$, $x_a$ and $x$ have a non-monotonous temperature dependence and do not reach the 3D limit $x=5$ even at $T=10$~K, although the  phonon 2D-3D crossover temperature for a 100~nm slab is $T_C \approx 240$~mK~\cite{PhysRevB.70.125425.2004.Kuhn, PhysScr.94.105704.2019.Anghel}.
This is due to the fact that although the phonon gas in the 100~nm thick slab is quasi-3D at 10~K, the energy of an average phonon is much smaller than the energy difference between the 2D electronic bands, so in this temperature range the electrons are still scattered only within the same band, $n=n'$.
Therefore, the higher temperature range corresponds here to the heat power exchange between a collection of 2D electron gases, with $n\le n_F$, and a 3D phonon gas.
In this case, the exponent $x$ approaches four,  satisfying the ansatz $x = s+2$, but with $s$ being the smaller dimensionality of the two subsystems--in our case, $s=2$ is the dimensionality of the electron subsystem.

\section{Conclusions} \label{sec_conclusions}

We studied the heat exchange between electrons and phonons in a suspended geometry, where a Cu film of thickness $d=10$~nm is placed inside a dielectric SiN$_x$ membrane of thickness $L=100$~nm, forming a layered structure.
We focused on investigating on how the location of the metal film influences the power flow, and found that at low temperatures it can change significantly -- at 10~mK it changes by an order of magnitude.
At sub-Kelvin temperatures, this metal film location dependence arises only from the coupling to the antisymmetric Lamb phonon modes of the membrane, whereas the symmetric Lamb-modes give a constant, location independent contribution. Moreover, the contribution of the antisymmetric modes goes to zero, if the metal film is placed at the center of the membrane.
The physical reason for this is that--by definition--the displacement field in the antisymmetric Lamb-modes is zero in the middle plane of the membrane.

In the low temperature limit, the temperature dependence of the symmetric mode contribution is $P^{(0)}_s \propto T^4$, whereas for the antisymmetric mode, $P^{(0)}_a \propto T^{3.5}$.
Therefore, if the metal film is not close to the center of the membrane, at low enough temperatures  $P_a$ prevails over $P_s$ and the total heat power flux has the simple temperature dependence $P(T_e, T_{ph}) \propto T_e^{3.5} - T_{ph}^{3.5}$.
In the opposite case, the symmetric mode dominates and $P(T_e, T_{ph}) \propto T_e^4 - T_{ph}^4$.
A consequence of this is that electrons and phonons can be much more efficiently decoupled at low temperatures by placing the metallic film in the center of the membrane.
This may also help considerably for electron cooling and noise reduction in ultrasensitive nanosensors.

In a wider temperature range, the exponent $x$ of the temperature dependence has a complicated, non-monotonous dependence on the temperature and on the metal film location.
For the antisymmetric mode, it varies from $\sim 3.5$ to $\sim 7$, whereas for the symmetric mode, it varies from from $\sim 4$ to $\sim 5.7$.
The bulk 3D limit, corresponding to $x=5$, was not achieved even at $T=10$~K, due to the high energy difference between the 2D electronic bands, but instead, the limit of $x=4$ is approached at $T=10$~K.

\section{Acknowledgments}

D.V.A. and M.D. acknowledge financial support by the Ministry of Education, UEFISCDI projects PN23210101 and PN23210204.
I.J.M. acknowledges support by the Academy of Finland project number 341823.

\end{document}